\documentclass{vow2008}

\title{Multi-wavelength data handling in current and future surveys: the possible role of Virtual Observatory}
\author[1,3]{C. Vignali}
\author[2]{F. Fiore}
\author[3]{A. Comastri}
\author[4]{M. Brusa}
\author[3]{R. Gilli}
\author[4]{N. Cappelluti}
\author[5]{F. Civano}
\author[3]{G. Zamorani}
\affil[1]{Dipartimento di Astronomia, Universit\`a degli Studi di Bologna, Italy}
\affil[2]{INAF -- Osservatorio Astronomico di Roma, Monteporzio, Italy}
\affil[3]{INAF -- Osservatorio Astronomico di Bologna, Italy}
\affil[4]{Max-Planck-Institut f\"ur Extraterrestrische Physik (MPE), Garching, Germany}
\affil[5]{Harvard-Smithsonian Center for Astrophysics, Cambridge MA, USA}

\usepackage{graphicx}
\newcommand{\ltsima}{$\; \buildrel < \over \sim \;$}
\newcommand{\simlt}{\lower.5ex\hbox{\ltsima}}
\newcommand{\gtsima}{$\; \buildrel > \over \sim \;$}
\newcommand{\simgt}{\lower.5ex\hbox{\gtsima}}

\newcommand{\cgs}{ ${\rm erg~cm}^{-2}~{\rm s}^{-1}$}

\def\lesssim{\mathrel{\hbox{\rlap{\hbox{\lower4pt\hbox{$\sim$}}}\hbox{$<$}}}}
\def\gtrsim{\mathrel{\hbox{\rlap{\hbox{\lower4pt\hbox{$\sim$}}}\hbox{$>$}}}}

\def\arcsec{\hbox{$^{\prime\prime}$}}
\def\micron{\hbox{$\mu$m}}

\def\ab1450{$AB_{1450(1+z)}$}

\def\xray{\hbox{X-ray}}

\def\sfr{M$_{\odot}$~yr$^{-1}$}
\def\edd_ratio{$\log\ L_{\rm bol}/L_{\rm Edd}$}

\def\chandra{{\it Chandra\/}}

\def\heao1{{\it HEAO-1\/}}
\def\hst{{\it {\it HST}\/}}

\def\spitzer{{\it Spitzer\/}}
\def\herschel{{\it Herschel\/}}

\def\rosat{{\it ROSAT\/}}

\def\xmm{{XMM-{\it Newton\/}}}

\def\erosita{{\it eROSITA\/}}
\def\wise{{\it WISE\/}}
\def\vista{{VISTA\/}}
\def\ps{{Pan-STARRS\/}}
\def\lsst{{LSST\/}}

\begin{document}


\maketitle

\begin{abstract}
Here we review some of the main issues related to multi-wavelength source identification 
and characterization, with particular emphasis on the field 
of \xray\ surveys carried out over the last years. 
This complex and time-consuming process is going to represent one of the main difficulties 
over the coming years, when significantly larger surveys, both in area and depth, 
will be carried out with the new generations of space- and ground-based facilities like e.g. 
\erosita, \wise, \vista, \ps, and \lsst. 
The Virtual Observatory can offer a reliable way to approach to a new concept of data handling and 
multi-wavelength source characterization, provided that uniform and rigorous data analyses and 
extensive quality checks are performed. 
\end{abstract}

\section{Introduction}
\label{introduction}
The increasing number of large surveys carried out over the last decade has revolutionized the 
fields of observational cosmology and astrophysics but, at the same time, it has posed several problems.
First of all, despite former surveys, which were primarily focused on censing and determining the 
properties of a given source class in a given band, recent surveys have been truly characterized by 
an observational multi-wavelength approach, whose final goal is to estimate the physical 
source properties over the entire electromagnetic spectrum and investigate how different selection 
criteria at different wavelengths are able to provide a complete census of such sources. 
Although an unbiased source selection is hard to achieve, current surveys in e.g. the obscured 
Active Galactic Nuclei (AGN) research field have allowed us to assess how biases work in source selection at 
different wavelengths, e.g. in the mid-IR vs. X-rays 
(e.g., Fiore et al. 2008, 2009; Donley et al. 2008 
and references therein). Similar conclusions can be easily extended to other source classes and bands. 

If, on the one hand, this approach is clearly consuming in terms of telescope/facility time and man-power, on the 
other hand it allows for a proper characterization of the intrinsic properties of a given source population. 

%
Although the science goals which motivate surveys at different wavelengths may be different, 
there are some common aspects which will be discussed in the following in order to understand 
the complex processes behind the build-up of source catalogs and the achievement of the 
published results.

\section{Source association and identification: 
examples from the \xray\ surveys in the COSMOS field}
\label{xray_surveys}
Motivated by the experience gained over the last decade in the field of \xray\ surveys 
by members of our collaboration, hereafter some practical examples and problems related to 
the process of \xray\ source identification will be presented. In particular, we will mention 
some issues concerning the recent surveys carried out with \xmm\ and \chandra\ in the 
Cosmic Evolution Survey (COSMOS) field, 
keeping in mind that some 
problems can be generalized to other surveys and wavelengths. 

The COSMOS survey\footnote{http://cosmos.astro.caltech.edu.} (Scoville et al. 2007a), 
an \hst\ Treasury Project, is a deep and wide 
extra-galactic survey over a contiguous $\approx$~2~deg$^{2}$ field 
designed to have sufficient area to limit the effects of cosmic variance in the 
investigation of e.g. large-scale structures (LSS; Scoville et
al. 2007b; Gilli et al. 2009), 
and depth to study the high-redshift \hbox{($z$$\approx$1--3)} galaxy and AGN populations. 
Among the wealth of multi-wavelength data (not all covering the entire area; 
see the review talk by McCracken, these proceedings) in this region of the sky, 
the \xray\ observations carried out with \xmm\ (1.55~Ms) and \chandra\ (1.8~Ms) -- 
the so-called XMM-COSMOS and C-COSMOS surveys (Hasinger et al. 2007; Elvis et al. 2009) --  
cover the entire 2~deg$^{2}$ of the COSMOS field and the innermost 0.9~deg$^{2}$, respectively. 
While \xmm\ observations are able to achieve a \hbox{0.5--2~keV} \hbox{(2--10~keV)} 
flux limit over 90\% of the area of $\approx1.7\times10^{-15}$~\cgs\  
($\approx9.3\times10^{-15}$~\cgs; Cappelluti et al. 2009), 
making \xray\ spectral analysis for a significant fraction of sources feasible (e.g., Mainieri et al. 2007), 
\chandra\ allows for a factor $\approx$~10 deeper X-ray flux detections in both the soft and hard bands. 
These depths are well suited for both AGN and normal/starburst galaxy studies in the \xray\ band 
(Ranalli et al., in preparation) and, given the typically low background in \chandra\ exposures, 
allow detection of faint \xray\ emission using stacking analysis techniques (see Fiore et al. 2009 
for a recent application). 
The final XMM- and C-COSMOS catalogs comprise 1822 and 
1761 point-like \xray\ sources above a detection likelihood of 10 and 10.8, respectively 
(Cappelluti et al. 2009; Elvis et al. 2009).

\subsection{Matching source catalogs}
\label{xray_matches}
Despite the ultra-deep exposures in the \chandra\ deep fields (North and South; see 
Hasinger \& Brandt 2005 for a review, and Luo et al. 2008 for recent updates), 
which are focused on observing the same portion 
of the sky (i.e., centered on the same aimpoint) repetitively, the mosaics in most of 
the current \xray\ surveys are based on observations of neighbouring regions with large overlaps 
among the different observations (but, for a different observing strategy, 
see AEGIS; Laird et al. 2009). 
Both XMM-COSMOS and C-COSMOS surveys are based on a tiling scheme which, in the latter 
survey, allows us to achieve a remarkably uniform exposure over the inner 0.45~deg$^{2}$ 
(for details, see Elvis et al. 2009 and Puccetti et al., in preparation). 
This translates directly into a sharply defined flux limit. 
In such a complex tiling strategy the same source is detected in more than one pointing 
with largely different Point Spread Functions (PSFs), 
possibly causing a worse source position reconstruction, hence a larger positional error in the final catalog, 
despite the potentially sharp \chandra\ PSF (FWHM$\approx$~0.5\arcsec) for on-axis source detection. 

The first drawback of having a few arcsec ``error box'' associated to each \xray\ source, depending on the 
source position across the field-of-view and the observed number of counts, is related to the match of 
the \xray\ sources to the counterparts observed at other wavelengths. This has strong 
consequences in the process of source identification and classification via optical and near-IR 
spectroscopy, making the already difficult task of spectroscopy (because of the faint magnitudes 
of many counterparts) quite challenging, since a larger number of sources should be observed and identified. 

Having these issues in mind, the \xray\ source catalogs in both XMM-COSMOS and C-COSMOS surveys 
have been matched to catalogs at other wavelengths using the deep and good-quality 
I, K, and 3.6~\micron\ data (Brusa et al. 2007; Brusa et al., in preparation; Civano et al., in preparation). 
In particular, using near-IR data allows for a more complete source association, since a fraction of \xray\ sources, 
especially at the very faint flux limits probed by \chandra, comprises obscured, optically faint (because 
obscured and at high redshift) AGN, as expected from the AGN synthesis models of \xray\ background (XRB; e.g., 
Gilli, Comastri \& Hasinger 2007). 

The adopted source matching method consists of the extensively used 
likelihood ratio (LR) technique (Sutherland \& Saunders 1992; for some recent applications of 
this method, see Ciliegi et al. 2003; Brusa et al. 2005; Laird et al. 2009; Motch, these proceedings). 
For a given optical/near-IR counterpart, LR is defined as the ratio between the probability that the 
source is the correct identification and the corresponding probability for a background, unrelated object. 
The chosen threshold for LR should therefore be a a careful compromise between completeness (i.e., most of the sources 
are finally recovered) and reliability (i.e., the number of possible spurious associations must be 
kept as low as possible). 
Without entering in too many details about this procedure, 
we note that this method has produced reliable source 
associations for the \xray\ sources in both XMM- and C-COSMOS surveys. 
For instance, $\approx$~80\% of the XMM-COSMOS sources had already a counterpart using the I and K-band catalogs, 
and this fraction increased by $\approx$~14\% using \chandra\ sharper positions. 
The \spitzer\ IRAC 3.6~\micron\ data increased the number of associations 
by a further 9\% and 5\% for the XMM- and C-COSMOS sources, respectively. 
This means that the final fraction of \xray\ sources without an optical/near-IR match (i.e., cases where either 
there are multiple sources in the \xray\ ``error box'' above the adopted LR threshold 
or at least one possible counterpart below the LR threshold) 
is 11\% in XMM-COSMOS (Brusa et al., in preparation) and only 1\% in C-COSMOS (Civano et al., in preparation), 
thus allowing for statistically reliable studies of the multi-wavelength properties of the \xray\ sources, 
including those with faint optical counterparts. 

Besides the overall good matching results, 
we note that the adoption of the LR technique requires significant efforts, 
mostly aimed at verifying the reliability of source associations and identifications beyond 
their ``statistical'' validity. 
The complexities and subtleties of such procedure have been faced -- at least in the COSMOS survey -- using 
visual inspection of the candidate counterparts on a source-by-source basis, requiring significant man-power 
on month timescales, 
given the large number of sources available and the continuous updates in the reference catalogs, 
which are extensively used before their final releases.  

In this context, Virtual Observatory (VO) can be of primary interest for people working in 
the field of galactic and extra-galactic surveys, provided that the available multi-wavelength catalogs 
are as uniform as possible (e.g., for photometric-redshift analyses and spectral energy distribution studies, 
magnitudes within the same extraction region should be homogeneously selected) and rigorously checked. 
If such conditions were satisfied, a further implementation to current VO matching tools could 
be the selection of one or more matching methods within VO itself, and running tests of source reliability, 
possibly using simulations to reassure about the obtained results. 
Given the way people are organized in present surveys (i.e., each group is mostly involved in 
reducing and analyzing data at a given wavelength, and finally in building up a source catalog), 
it must be said that implementing source-matching tasks in VO could be of limited use. 
On the other hand, given the large amount of data that will be available 
in the near future using the next generation of 
ground- and space-based facilities, this issue will become more and more important and time-consuming, 
thus invoking for a significantly different approach in the way we currently deal with observational data.

\section{Coming soon: a quick look at some of the new facilities} 
\label{coming_next}
In the following, some of the main next-generation facilities in the survey field will be outlined, along 
with potential problems in data handling and future catalog construction. 
These facilities will provide powerful tools for source identification on relatively short 
timescales. 

\subsection{\erosita}
\label{erosita}
Over the next few years, one of the major advances in the field of \xray\ surveys will be provided by 
the \erosita\ mission (Predehl et al. 2007) which is scheduled for launch in 2012.\footnote{http://www.mpe.mpg.de/projects.html\#erosita.} 
This mission is focused on \xray\ surveys, being the natural continuation of the former 
all-sky survey performed in the soft band by the \rosat\ satellite in early 90' (RASS; Voges et al. 1999). 
With an estimated PSF Half Energy Width (HEW) of $\approx$~15\arcsec\ for on-axis observations 
(and average PSF HEW of $\approx$~25\arcsec over the entire field-of-view of 0.83~deg$^{2}$), 
\erosita\ is primarily intended to perform a 4-yr all sky survey, reaching 
a flux limit of $\approx1.0\times10^{-14}$ 
($\approx1.2\times10^{-13}$)~\cgs\ in the \hbox{0.5--2~keV} \hbox{(2--10~keV)} band, 
with typical exposures of $\approx$~2~ks, and a 0.5-yr deep survey over 400~deg$^{2}$ 
at flux limits about a factor $\approx$~5 and $\approx$~13 fainter in the two bands, respectively 
(with an average exposure of $\approx$~27~ks). 

For what concerns the study of the AGN physics and evolution, the planned all-sky survey will allow for 
detections of $\approx$~4~million AGN, 25000 (2100) of which being above redshifts $z$=3 (4). 
These numbers derive from the most up-to-date XRB synthesis models (Gilli et al. 2007) and calibrations from the \erosita\ team; 
in particular, the on-flight background (cosmic plus particle) has been simulated for a L2 orbit. 
In survey mode, where source statistics is photon limited, 
the sensitivity has been estimated by imposing that a source can be detected if at least 9 net counts 
are present and by assuming a tipical power-law spectrum with photon index $\Gamma$=1.8 and Galactic column density. 
Then the source counts have been converted into a flux using the response matrices available at the time of writing. 
In deep survey mode, the background counts have been accumulated in 30\arcsec\ circular apertures, 
and limiting counts have been estimated by imposing to have a 3$\sigma$ detection above the background. 
Counts were then converted into fluxes by using the same \xray\ spectral model as above.

This approach will represent a major improvement in the field of statistical studies of AGN; 
for comparison, less than one thousand AGN are detected in the \xray\ ultra-deep exposures in the CDF-N and CDF-S, 
and of the order of a couple of thousands in COSMOS (see $\S$\ref{xray_matches}). 
The \erosita\ deep survey, although on a limited region, 
will allow for detections of \xray\ faint AGN at high redshifts, which are currently poorly sampled 
by current \xray\ surveys (e.g., see Fig.~3 of Vignali et al. 2005).  

The large amount of sources detected by \erosita\ will require, at some level, 
multiple campaigns of photometric and spectroscopic follow-up observations. 
Just to give an idea of the numbers involved in these follow-ups, at the conservative 
\hbox{0.5--2~keV} flux limit of \hbox{$\approx4\times10^{-15}$}~\cgs\ for the \erosita\ deep survey, 
we find $\approx$~570 sources in XMM-COSMOS, with the faintest sources having an I-band (3.6~\micron) 
magnitude of $\approx$~24 (22; both magnitudes are in the AB system). 
Covering 400~deg$^{2}$ at such depth is currently hard to accomplish, 
and the whole spectroscopic coverage will be even more complex and time-consuming. 
Given the \erosita\ PSF, the source identification will not be an easy process 
and will take advantage of follow-up observations, especially in the near-IR, where 
the planned \wise\ and \vista\ surveys will provide a significant contribution to 
source identification, hence to observational cosmology. 
These and other planned facilities that will be available to the 
community over the coming years are briefly reviewed in the following. 

\subsection{\wise}
\label{wise}
The Wide-field Infrared Survey Explorer (\wise) is a NASA mission\footnote{http://wise.ssl.berkeley.edu.} (launch scheduled for mid 2009) aimed at surveying the entire 
sky in four bands (3.3, 4.7, 12, and 23~\micron), with sensitivities of 0.12, 0.16, 0.65 and 2.6~mJy (after eight repetitions of 
8.8s each), respectively. Such sensitivities are worse than those achieved by the instruments onboard the \spitzer\ satellite, 
but are a factor $\approx$~1000 better than those obtained with the IRAS surveys (Mainzer et al. 2006). 
Furthermore, although the resolutions allowed by \wise\ are typically worse (FWHM of 6\arcsec\ and 12\arcsec\ in the near-IR and mid-IR, 
respectively) 
than \spitzer, the large field-of-view of \wise\ (47~arcmin$^{2}$) will allow to perform a survey of the entire sky in just six months. 
This means that studies like those regarding obscured high-redshift AGN by means of mid-IR--to--optical flux ratio selection, 
as those carried out recently by several authors (e.g., Daddi et al. 2007; Fiore et al. 2008, 2009; Dey et al. 2008; Lanzuisi et al. 2009), 
will be possible over much larger sky regions (although at higher flux limits), provided that coverage at other wavelengths (to select 
and properly characterize the sources) will be granted. 

\begin{figure}
\includegraphics[width=0.36\textwidth,angle=-90]{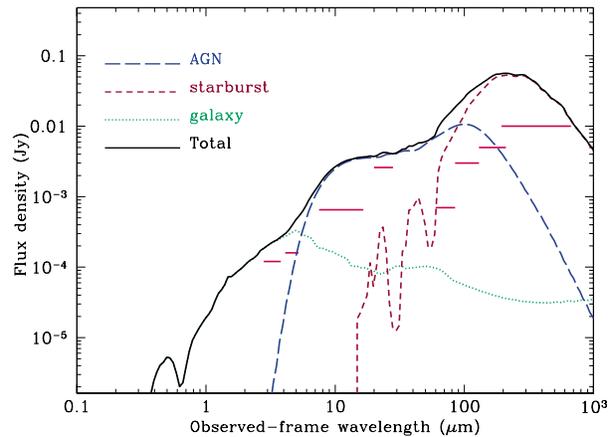}
\caption{Observed-frame SED of an \xray\ selected, luminous and obscured quasar from the 
HELLAS2XMM survey (Vignali et al. 2009) at $z$=1.957. 
The ``total'' SED (solid line) is the summed contribution of 
a stellar component (dotted line), a torus component (long-dashed line), and a starburst component (short-dashed line). 
The sensitivity and confusion limits expected for \wise\ (four near-IR and mid-IR bands) and 
\herschel\ (three mid-IR/far-IR bands with PACS and three far-IR bands extending from $\approx$~200 to 670~\micron\ with SPIRE) 
are shown as horizontal bars, whose length is indicative of the band covered by each filter or instrument 
(as in the case of SPIRE).}
\label{scuba_SED_persp}
\end{figure}

Similarly, detection of obscured AGN from shallow and moderate-depth \xray\ surveys up to high-redshifts will be possible, 
as shown in Fig.~\ref{scuba_SED_persp}, where the optical, infrared and sub-mm spectral energy distribution (SED) 
of a luminous Type~2 quasar at $z$=1.957 detected by the HELLAS2XMM survey (Fiore et al. 2003) is presented. 
The intense star formation derived for this source ($\approx$~1500~\sfr) is suggestive of an AGN passing through a complex phase, 
being caught presumably before the process of expelling the obscuring gas (thus quenching the star formation) 
and becoming a broad-line unobscured quasar has started. AGN with similar properties do not seem to be uncommon at high redshifts, 
as shown in recent work (e.g., Mainieri et al. 2005; Polletta et al. 2008). 
Through proper broad-band SED fitting, \wise\ and \herschel\ will be able to constrain the individual components 
(stellar, AGN, and starburst) and estimate the nuclear bolometric luminosity (see Vignali et al. 2009 for details). 

\subsection{\vista}
\label{vista}
The Visible and Infrared Survey Telescope for Astronomy (\vista)\footnote{http://www.vista.ac.uk/index.html.} 
is a 4-m wide-field (1.65~deg diameter) survey telescope equipped with a near-IR camera facility 
(at least at the beginning of the operations, although it is also capable of operating in the visible with a suitable camera), 
covering the 0.85--2.3~\micron\ wavalength range with five broad-band filters ($Z$, $Y$, $J$, $H$, and $K_{\rm s}$) and a narrow-band filter 
centered at 1.18~\micron. 
Located at the ESO Cerro Paranal Observatory in Chile, the \vista\ planned capabilities are such to suitably match the needs 
of the 8~m-class telescopes of VLT. The large field-of-view betrays that \vista\ is mainly designed for carrying out 
large-area surveys in the near-IR domain. In this regard, it is properly suited for follow-up identification programs 
(e.g., see $\S$~\ref{erosita}). 
One of the proposed surveys is the \vista\ Hemisphere Survey that will image $\approx$~20000~deg$^{2}$ 
in the $J$ and $K_{\rm s}$ bands. 
About 40\% of this survey, dedicated to a deep observational program 
searching for the dark energy (Dark Energy Survey -- DES), will cover $\approx$~5000~deg$^{2}$ also in the $H$ band over initially 125 nights down to 
10$\sigma$ AB magnitude limits in the $J$, $H$, and $K_{\rm s}$ bands of 20.4, 20.0, and 19.4, respectively (Banerji et al. 2008). 
Furthermore, the UltraVISTA survey is aimed at imaging the COSMOS field down to unprecedented depths, 
taking advantage also of the deep 3.6--4.5~\micron\ coverage with the planned \spitzer\ Extended Deep Survey (SEDS) ``warm-mission'' project. 
The UltraVISTA survey will use all of the near-IR filters, 
including the narrow-band one to search for Ly$\alpha$ emitters at very high redshifts ($z\approx8.8$). 
Finally, we mention also the \vista\ Deep Extragalactic Observations Survey (VIDEO), 
which will cover 15~deg$^{2}$ of the sky (some of which already included in SWIRE) 
with the $Z$, $Y$, $J$, $H$ and $K_{\rm s}$ filters, jointly with the recently approved \spitzer\ Extragalactic Representative Volume Survey (SERVS) project.

\subsection{\ps}
\label{panstarrs}
The Panoramic Survey Telescope and Rapid Response System (\ps) consists of four telescopes of 1.8m each located in Hawaii 
(the prototype PS1 being already on site).\footnote{http://pan-starrs.ifa.hawaii.edu/public.} Its innovative 
design for wide-field imaging, which comprises a 64$\times$64 CCD array (600$\times$600 pixel each, for a total of 1.4 billion pixels), 
is such to allow observations of the available sky ($\approx$~30000~deg$^{2}$) several times each month, 
down to $r\approx24.0$ (5$\sigma$). 
Although the immediate goals of \ps\ are related to discover and characterize Earth-approaching objects, the huge amount of images guaranteed by 
this facility will allow for fast photometric observations of large sky regions (up to $\approx$~3000-6000~deg$^{2}$ in one night). 
Furthermore, the wavelength range of the available filters ($g$, $r$, $i$, $z$, and $y$, covering the $\approx$~4700--10500~\AA\ interval) 
is suited to provide a first-order characterization of the sources, including an estimate of their redshift via photo-$z$ techniques. 

\subsection{\lsst}
\label{lsst}
The Large Synoptic Survey Telescope (\lsst)\footnote{http://www.lsst.org/lsst.} 
represents one of the most revolutionary ideas for the next decade. 
Aimed at probing dark energy and dark matter, mapping the Solar system and the Milky Way, and exploring transient object, \lsst\ will offer 
unique opportunities for large-area surveys. It consists of a large, wide-field ground-based system 
designed to obtain multiple images covering the $\approx$~30000~deg$^{2}$ of sky which are visible from 
Cerro Pach\'{o}n in Chile. The current baseline design is characterized by an 8.4m 
(6.5m effective) primary mirror with a 9.6~deg$^{2}$ field-of-view and a 3.2-billion pixel camera (Gilmore et al. 2008), 
and allows observations of about 10000~deg$^{2}$ of sky using pairs of 15s exposures in two of the six 
photometric bands ($u$, $g$, $r$, $i$, $z$, and $y$) every three nights on average, 
with typical 5$\sigma$ depth for point-like sources of $r\approx24.7$ (Ivezic et al. 2008). The ``deep-wide-fast'' survey mode will 
be assigned 90\% of the time, thus allowing for observations of $\approx$~20000~deg$^2$  in the six bands (over the ten years of operations) 
and reaching a final depth of $r\approx27.7$. 
The remaining 10\% of the time will be allocated to special programs such as the ``Very Deep and Fast time domain survey'', reaching 
$r\approx28$. 
Three billion galaxies are expected to have reliable photometric redshifts at the end of the observations, and 20--80 million AGN 
should be detected (i.e., two orders of magnitude larger than the current AGN census); a fraction of AGN 
(depending on their absolute magnitudes, redshifts, and observed time intervals) 
will be recognized as such on the basis of their variability, thanks to the multiple observing-epoch strategy. 

\subsection{Spectroscopic facilities: HETDEX and SDSS-III}
\label{spec_facilities}
While the previous facilities will focus on imaging observations of large sky regions, 
here we present a couple of new facilities which will be available to the community to carry out 
spectroscopic programs. The first is the HETDEX (Hobby-Eberly Telescope Dark Energy Experiment)/VIRUS (Visible Integrated-Field Replicable Unit 
Spectrographs) project,\footnote{http://hetdex.org/hetdex/virus.html.}
which consists on an innovative set of spectrographs (for a total number of $\approx$~34000 fibers) 
to observe thousands of galaxies each night. 
The HETDEX/VIRUS survey, with the primary goal of exploring the issues related to the dark energy through a three-dimensional 
map of a large volume of space, is scheduled for 2010--2013 and is going to provide redshifts for about 1~million galaxies. 

Over the 2009--2014 period, the Sloan Digital Sky Survey III (SDSS-III)\footnote{http://www.sdss3.org.} 
will carry out a program of four surveys on three scientific themes: 
dark energy, dynamics and chemical evolution of the Milky Way, and studies of planetary systems. 
In particular, the Baryon Oscillation Spectroscopic Survey (BOSS) project, in order to map the spatial distribution of luminous galaxies and quasars, 
will survey 10000~deg$^{2}$ and provide redshifts for $\approx$~1.5 million galaxies up to $z$=0.7, 
and 160000 quasars at $z$$\approx$2.2--3, using a 1000-fiber spectrograph with resolution R$\approx$2000.

\section{The possible role of the virtual Observatory}
Given the large amount of data that will be available to the community in the future, 
it seems plausible that VO can have an effective role in data handling and related tasks, as actually 
presented by many authors during this workshop. 
As pointed out in $\S$\ref{xray_matches}, large multi-wavelength datasets require homogeneous 
data selection and validation, and quality checks. 
Having said this, it is imperative for all of the people releasing reduced data and corresponding source catalogs to 
carry out high-level quality checks and exhaustive accompanying information. 

From an \xray\ perspective, we have shown that the process of 
building up and cross-correlating catalogs using proper matching criteria (e.g., the likelihood ratio technique) is time-consuming 
and requires, at some level, visual inspection to support the statistical reliability of the matched sources. 
If, on the one hand, some checks are clearly beyond the possibilities of VO, on the other hand 
data homogenization and automatization of some matching procedures (and corresponding checks) would be 
highly appreciated by the community. Such implementations would allow for a quicker, more extensive and exhaustive exploitation 
of the available datasets. 
Similarly, automated tasks to perform stacking analyses at different wavelengths and estimate photometric 
redshifts using broad-band matched catalogs would greatly enhance the use of VO, especially by people not directly (or deeply) 
involved in all the issues and problems related to surveys and data handling. 

Synergy with VO is undoubtedly a challenging issue over the coming years 
when billion-object catalogs will be available and 
the necessity of optimizing the many-sided scientific outputs on relatively 
short timescales will become more and more pressing.

\section*{Acknowledgments}
CV acknowledges ESO financial support; the authors would like to thank 
the members of the HELLAS2XMM, ELAIS-S1, XMM-COSMOS, and 
C-COSMOS collaborations. 


\end{document}